\documentclass[a4paper,12pt]{article}

\usepackage{sint} 
\newcommand{\csw}{c_\text{\tiny SW}}
\newcommand{\ca}{c_\text{A}}
\newcommand{\fa}{f_\text{A}}
\newcommand{\fp}{f_\text{P}}

\usepackage{subfig}
\usepackage{booktabs}
\usepackage{amsmath}
\usepackage{amssymb,bbm}
\usepackage{graphicx}
\usepackage{multirow}

\begin{document} 

\begin{titlepage}

\begin{flushright}
\small{
CERN-PH-TH/2013-082\\
TCD-MATH 13-06\\
}
\end{flushright}
\vskip5mm

\begin{center}
{\Large\bf
Improvement of $N_f = 3$ lattice QCD with Wilson
   fermions and tree-level improved gauge action
}
\end{center}

\vskip 0.1cm
\begin{center}
John Bulava$^{\scriptscriptstyle a,b}$ and
Stefan Schaefer$^{\scriptscriptstyle a}$\\
{
\vskip 2.0ex
$^{\scriptstyle a}$
CERN, Physics Department, 1211~Geneva~23, Switzerland 
\vskip 2.0ex
$^{\scriptstyle b}$
School of Mathematics, Trinity College, Dublin 2, Ireland
\vskip 2.0ex
}
\vskip 0.5cm
{\bf Abstract}
\vskip 0.1ex
\end{center}

We determine the parameter $\csw$ required for 
$\mathrm{O}(a)$-improvement of the three flavor Wilson fermion action 
together with the tree-level Symanzik improved gauge action. The 
standard improvement condition is employed for a range of couplings. 
Additionally, we perform  a
check of the volume independence of $\csw$ and provide a preliminary estimate 
of the lattice spacing at our largest values of $g_0^2$.


\vskip 2.0ex

\vfill
\eject

\end{titlepage}


\section{Introduction \label{sec:intro}}
The continuum limit is an essential part of lattice QCD calculations.
In order to control this extrapolation in the lattice spacing $a$, it
is advisable to work with a discretization whose leading cut-off effects
are $\mathrm{O}(a^2)$. The theory and reduction of
scaling violations in on-shell quantities is
well-established~\cite{impr:Sym1,impr:Sym2,impr:onshell} and for Wilson
fermions~\cite{Wilson} $\mathrm{O}(a)$ improvement can be achieved by
adding a single dimension-five operator to the action~\cite{impr:SW}.
This requires the non-perturbative tuning of
its coefficient $\csw$ and can be performed
using a standard procedure~\cite{Luscher:1996ug,impr:csw_nf2}. 
Simulations with the resulting action in two-flavor flavor 
QCD have exhibited the expected moderate scaling
violations~\cite{DellaMorte:2008ad}.

Here we present a determination of the coefficient $\csw$ using the
standard procedure for $N_\mathrm{f}=3$ flavor QCD with the tree-level
improved L\"uscher--Weisz gauge action~\cite{Luscher:1984xn}. We choose
this gauge action because it has been demonstrated to possess superior
scaling properties in pure gauge theory~\cite{silvia:universality}.

This paper is organized as follows. After defining our lattice regularization 
in Sec.~\ref{sec:defs}, we summarize the standard improvement programme in
Sec.~\ref{sec:imp}.  In Sec.~\ref{sec:sims} details of the numerical
computations are given together with the resultant interpolating formula
for the improvement coefficient $\csw(g_0^2)$. We also give 
results for a renormalized quantity which suggests that the range of
bare couplings covered by our simulations extends at least to a lattice
spacing of $a\approx0.09\,\mathrm{fm}$.

\section{Lattice setup\label{sec:defs}}
The $\mathrm{O}(a)$-improved Wilson Dirac operator~\cite{Wilson,impr:SW}
is given by
\begin{equation} D_\text{W} = 
\frac{1}{2}\sum_{\mu=0}^3 \{\gamma_\mu(\nabla^*_\mu+\nabla_\mu) -\nabla^*_\mu\nabla_\mu\}
+\csw\sum_{\mu,\nu=0}^3\frac{i}{4}\sigma_{\mu\nu}\widehat F_{\mu\nu}+m_0
\end{equation} 
with $\nabla_\mu$ and $\nabla_\mu^*$ the covariant
forward and backward derivatives, respectively, and $\widehat F_{\mu\nu}$
the standard discretization of the field strength
tensor~\cite{impr:pap1}.  The bare mass $m_0$ will be replaced below by
the hopping parameter $\kappa$ with $m_0=(\kappa^{-1}-8)/2$.

The gauge action $S_G$ contains sums over all oriented $1\times1$ plaquettes
as well as all
$1\times2$ rectangles, which are denoted by the sets 
$\mathcal{S}_0$ and $\mathcal{S}_1$, respectively,
\begin{align}
S_G = \beta\, \sum_{k = 0,1} c_k \, \sum_{\mathcal{C} \in \mathcal{S}_k} 
w_k(\mathcal{C})\; \mathrm{tr} \{ 1 - U(\mathcal{C})\},
\end{align}
where $\beta=6/g_0^2$, $c_0 =
5/3$ and $c_1=-1/12$. The weight factor
$w_k(\mathcal{C})$ is set to unity for all loops away 
from the boundaries.
Schr\"{o}dinger functional~\cite{SF:LNWW,SF:stefan1} boundary
conditions are imposed on the gauge fields such that 
\begin{align}\label{eq:gbc}
w_0(\mathcal{C}) &= \begin{cases} \frac{1}{2}, & \text{all links in $\mathcal{C}$ are on a boundary}  
\\1, & \text{otherwise}  
\end{cases}
      \intertext{and}
w_1(\mathcal{C}) &= \begin{cases} \frac{1}{2}, & \text{all links in $\mathcal{C}$ are on a boundary}  
\\\frac{3}{2}, &  \text{$\mathcal{C}$ has exactly two links on a boundary} 
\\1, & \text{otherwise.}  \end{cases}
\end{align}
This corresponds to `Choice B' of Ref.~\cite{Aoki:1998qd} and 
guarantees boundary $\mathrm{O}(a)$ improvement at tree-level of perturbation 
theory. We also set the fermionic boundary counter term according to the
1-loop formula~\cite{Aoki:1998qd} 
\begin{align}
c_F&=1-0.0122 \ C_F\,  g_0^2, & \text{with}\ & C_F=4/3 \,.
\end{align}
Finally, the values of the spatial links at the boundaries are fixed to
\begin{align}\label{e:bnd}
U(x, k)|_{x_0 = 0 } &= \exp\{aC_k\} \; , &
C_k &= \frac{i}{6L}\, \mathrm{diag}(-\pi, 0, \pi) \, ,
\\
U(x, k)|_{x_0 = T} &= \exp\{aC_k'\}\; , &
C_k' &= \frac{i}{6L}\, \mathrm{diag}(-5\pi, 2\pi ,3\pi)\, ,
\end{align}
while the fermion fields satisfy periodic boundary conditions in the 
spatial directions and the standard Schr\"odinger functional boundary
conditions in time.

\section{Improvement condition\label{sec:imp}}
The standard $\mathrm{O}(a)$ improvement programme relies on the PCAC relation, 
which involves the improved axial-vector current $(A_\text{I})_\mu^a$ and the 
pseudoscalar density $P^a$ given by 
\begin{align}\label{e:pcac}
(A_\text{I})_\mu^a&=A_\mu^a\, +\, a\, \ca\,
   \frac{1}{2}(\partial_\mu^*+\partial_\mu) P^a, && \\
     A^a_\mu(x) & = \bar \psi(x) \gamma_\mu\gamma_5
     \frac{\lambda^a}{2}\psi(x), &  P^a(x) & = \bar \psi(x)
   \gamma_5
        \frac{\lambda^a}{2}\psi(x) \,.
\end{align}
In the unimproved theory the unrenormalized PCAC relation
\begin{align}
\frac{1}{2} (\partial_\mu + \partial_\mu^*) \langle (A_I)_\mu^a(x)
   \mathcal{O} \rangle = 2 m \langle P^a(x) \mathcal{O} \rangle
   \label{e:pcacu}
\end{align}
is violated by terms of $\mathrm{O}(a)$. By using three different
choices of  $x$, $\mathcal{O}$ and ${\mathcal{O}'}$ one can define
$\csw$ and $\ca$, requiring that $m$ is the same in all three cases.
In this situation Eq.~\ref{e:pcac} holds  up to $\mathrm{O}(a^2)$.

This method has been applied to the $N_f=0,2,3,4$ cases for a variety of 
different
actions~\cite{Luscher:1996ug,impr:csw_nf2,impr:csw_nf3,Aoki:2005et,Tekin:2009kq,Cundy:2009yy}. 
For technical reasons, we use lattices with 
$T=2L-a$. However, this additional $\mathrm{O}(a)$ effect is irrelevant, as the
determination of $\csw$ using our improvement condition is ambiguous at 
$\mathrm{O}(a)$. We employ the PCAC 
relation with boundary operators $\cal O$ and $\cal O'$ on time slices
$x_0=0$ and $x_0=T$, respectively,
\begin{align}
\mathcal{O}^a&=a^6 \sum_{\mathbf{y},\mathbf{z}} \bar\zeta(\mathbf{y}) \gamma_5
\frac{\lambda^a}{2} \zeta(\mathbf{z})\,, &
\mathcal{O'}^a&=a^6 \sum_{\mathbf{y},\mathbf{z}} \bar\zeta'(\mathbf{y}) \gamma_5
\frac{\lambda^a}{2} \zeta'(\mathbf{z})\,.
\end{align}
The correlation functions which enter the PCAC
relation Eq.~\ref{e:pcac} are then 
\begin{align}\label{e:cor}
\fa(x_0) & = -\frac{1}{3}\langle A_0^a(x) \mathcal{O}^a \rangle\,,
&
\fp(x_0) & = -\frac{1}{3}\langle P^a(x) \mathcal{O}^a \rangle\,,\\
\fa'(T-x_0) & = +\frac{1}{3}\langle A_0^a(x) \mathcal{O'}^a \rangle\,,
&
\fp'(T-x_0) & = -\frac{1}{3}\langle P^a(x) \mathcal{O'}^a \rangle\,,
\end{align}
As has been suggested in Ref.~\cite{Luscher:1996ug}, effective masses
$M(x_0)$  defined by  
\begin{align}
M(x_0,y_0)= r(x_0)-s(x_0)\frac{r'(y_0)-r(y_0)}{s'(y_0)-s(y_0)}\,.
\label{e:M}
\end{align}
with
\begin{align}
r(x_0)&=\frac{1}{4}(\partial_0^*+\partial_0)
   f_\mathrm{A}(x_0)/f_\mathrm{P}(x_0)\quad \text{and} \quad
   s(x_0)=\frac{1}{2}a\partial_0^*\partial_0\,f_\mathrm{P}(x_0)/f_\mathrm{P}(x_0)
\end{align}
correspond to a particular choice of $\ca$ in the improved currents.
Since $M(x_0, y_0)$ renormalizes multiplicatively, it is a useful quantity to
define the quark mass at fixed $\beta$ and 
$\csw$. Specifically, we take as our definition of the quark mass 
\begin{equation}
M=\frac{1}{2}\big(M(L,L/2) + M(L-a,L/2) \big)\,.
\end{equation}
Tuning $M\approx0$ defines the values of $\kappa$ at which we impose 
the improvement condition. 

As discussed above, this improvement condition requires the
difference between two masses to vanish. In addition to $M(x_0, y_0)$, a second 
mass $M'(x_0, y_0)$ is considered where $r$ and $s$ in 
Eq.~\ref{e:M} are replaced by their primed counterparts. 
These  masses are evaluated at
$x_0=3T/4$ and $y_0=T/4$. However, because we have $T=2L-a$, we round the 
two arguments towards the center of the lattice. The parameter $\csw$ is
chosen such that the difference between these two masses $\Delta M$ is equal 
to its tree-level value $\Delta M^{(0)}$
\begin{equation}
\Delta M = M(3T/4,T/4)-M'(3T/4,T/4) \equiv \Delta M^{(0)} \,.
\label{e:imp}
\end{equation}

The numerical value of $\Delta M^{(0)}$ depends on the lattice geometry
and can be computed using the solution to the Dirac equation as given in
Sec. 6.2 of Ref.~\cite{impr:pap2}. For the $15\times8^3$ lattices
$a\Delta M^{(0)}=0.000393$. Additionally, this offset may be obtained
from measurements on free gauge fields. A stringent test of our entire
workflow is the reproduction of the $\Delta M^{(0)}$ obtained from
analytic calculations using simulations at large values of $\beta$.

In principle it would be preferable to keep the physical size of the
system $L$ constant as the continuum limit is approached. This, however,
turns out to be very costly in practice, because autocorrelations associated
with the topological charge sectors quickly become very large as
the lattice spacing is lowered.   We therefore opt to impose the
improvement condition at fixed $L/a$, where the contribution of
non-zero topological charge sectors decreases rapidly in the continuum
limit. In Sec.~\ref{s:fv} we will show that the effect of the finite
volume does not seem to be relevant at the current level of accuracy.

\section{Simulations}
\label{sec:sims}

For the simulations we use the \texttt{openQCD} code, 
which is publicly availible 
online\footnote{\texttt{http://luscher.web.cern.ch/luscher/openQCD}} and implements the lattice setup 
of Sec.~\ref{sec:defs} for several simulation algorithms. These
algorithms are the subject of
Ref.~\cite{Luscher:2012av} so we restrict ourselves to a
brief summary here.

We employ the HMC algorithm~\cite{Duane:1987de} with a  twisted-mass
Hasenbusch frequency splitting~\cite{algo:GHMC,algo:GHMC3} for a doublet
of two of the three degenerate quarks. For most ensembles with
$\beta\leq3.5$ twisted mass reweighting~\cite{Luscher:2008tw} is used,
i.e. we simulate with a small twisted mass $\mu=0.001$ and then
include a stochastically estimated reweighting factor to correct for
this in the measurement. This significantly increases the stability
of the simulation. The third quark is simulated using the RHMC
algorithm~\cite{algo:RHMC}. In all cases, we use a nine pole rational
approximation in the interval $[0.02,7.2]$ and correct for the
rational approximation with an additional stochastically estimated
reweighting factor.  All fermion determinants are factorized using
even-odd preconditioning. 

For the entire range of couplings, we generated $15\times8^3$ lattices
at a variety of $\csw$ values listed in Table~\ref{t:sims}.  In
addition, $23\times12^3$ ensembles at $\beta=3.8$ have been generated as
finite volume checks.  To obtain a preliminary estimate of the lattice
spacing at $\beta=3.3$ and $3.4$, we performed some $L=T$ runs (using a
modified version of the \texttt{openQCD} code) with $L/a=8$ and
zero boundary fields. These runs used  both our discretization and the
one of Ref.~\cite{Aoki:2009ix} and are summarized in
Table~\ref{t:scale}.  

For the $L/a=8$ lattices at the four smallest values of $\beta$ we use a
three-level hierarchical integration scheme in which the outermost level
employs the second order Omelyan-Mryglod-Folk (OMF) integrator, while
the two inner levels use the fourth order OMF
integrator~\cite{Omelyan2003272}.  The force from the pole closest to
the origin in the rational approximation is integrated on the coarsest
timescale, the remaining fermion forces on the intermediate timescale,
and the gauge force on the finest timescale. Four or five outermost
iterations together with a single iteration of the remaining two
integrators typically achieve $\approx90\%$ acceptance for molecular
dynamics trajectories of length $\tau = 2$.  
 
For the $L/a=12$ runs at $\beta=3.8$, the remaining $L/a=8$ runs and the
$T=L$ runs, a two level scheme with both levels using fourth order OMF
was found to be effective, with between 5 and 8 steps in the outermost
integrator and a single inner iteration.  

The integration schemes discussed above were very stable in all cases,
resulting in small Hamiltonian violations. The most difficult
simulations were those with $L/a=8,T=2L-a$ at $\beta=3.3$ at the
smallest value of $\csw$, but even those had only about $0.6\%$
trajectories with $\Delta H>10$. At $\csw=2.1$ this falls already to
$0.2\%$ and we had $0.08\%$ of such events at $\csw=2.4$.  At the
smallest $\csw=1.7$ for $\beta=3.4$ these trajectories occurred with
$0.1\%$, a percentage rapidly falling for larger values of $\csw$
and $\beta$.

\section{Results}
\label{sec:res}

We finally come to the results of our simulations. Our analysis strategy
is discussed in Sec.~\ref{s:ana} and results for $\csw(g_0^2)$ are
collected in Sec.~\ref{s:l8}, with a finite volume check in
Sec.~\ref{s:fv}. A preliminary scale determination is given in
Sec.~\ref{s:scale}. 

A summary of the ensembles generated for the determination of $\csw$
appears in Tab.~\ref{t:sims}, where we also give the total statistics
accumulated over several replica. These ensembles consist of $L/a=8$
lattices used for our final result as well as those used for the finite
volume check.  The ensembles generated for the preliminary scale setting
are discussed in Sec.~\ref{s:scale} and collected in Tab.~\ref{t:scale}.

\begin{table}

\centering
\small
\newcommand{\pp}{\hphantom{-}}
\begin{tabular}{@{\extracolsep{0.3cm}}llllllr}
\toprule
$\beta$ & $L/a$ & $\csw$  &$\kappa$  & $aM$          & $a\Delta M$   & MDU    \\
\midrule
\multirow{4}{*}{3.3}&\multirow{4}{*}{8}
 &$1.80  $ & $0.1434558$& $-0.0092(16)$ & $\pp0.0028(8)$ & $75622$ \\
&&$2.10  $ & $0.1372440$& $-0.0010(12)$ & $-0.0006(10)$ & $54336$ \\
&&$2.40  $ & $0.1315404$& $\pp0.0036(6)$ & $-0.0013(4)$ & $39002$ \\
&&$2.70  $ & $0.1266163$& $\pp0.0101(4)$ & $-0.0042(3)$ & $33104$ \\
\midrule
\multirow{4}{*}{3.4}&\multirow{4}{*}{8}
&$1.70$ &$0.1427535$  & $-0.0018(9)$ & $\pp0.0015(11)$ & $135310$ \\
&&$2.00$ &$0.1371025$  & $-0.0125(6)$ & $-0.0001(4)$ &    $115036$  \\
&&$2.30$ &$0.1318818$  & $-0.0089(5)$ & $-0.0021(3)$ &    $53282$ \\
&&$2.60$ &$0.1270442$ & $\pp0.0070(3)$  & $-0.0050(2)$ &  $63136$ \\
\midrule
\multirow{3}{*}{3.5}&\multirow{3}{*}{8}
      &$1.85  $ & $0.1374470$& $\pp0.0091(7)$ & $\pp0.0004(4)$ & $48860$ \\
&      &$2.20  $ & $0.1319060$& $\pp0.0010(3)$ & $-0.0020(2)$  & $62872$\\
&      &$2.55  $ & $0.1267940$& $\pp0.0109(4)$ & $-0.0056(3)$ & $37800$ \\
\midrule
\multirow{4}{*}{3.601}&\multirow{4}{*}{8}
&$1.50  $ & $0.1420500$& $-0.0009(14)$ & $\pp0.0030(7)$ &$19200$\\
&&$1.70  $ & $0.1387200$& $-0.0057(9)$ & $\pp0.0006(5)$ &$22200$\\
&&$1.90  $ & $0.1353560$& $-0.0010(6)$ & $-0.0002(4)$ &$20200$\\
&&$2.10  $ & $0.1319920$& $\pp0.0130(4)$ & $-0.0027(3)$ &$28600$\\
\midrule
\multirow{4}{*}{3.8}&\multirow{4}{*}{8}
&$1.20  $ & $0.1434300$& $\pp0.0134(10)$ & $\pp0.0039(5)$ & $47400$ \\
&&$1.40  $ & $0.1405500$& $\pp0.0041(7)$ & $\pp0.0023(5)$ &  $21942$ \\
&&$1.60  $ & $0.1376520$& $\pp0.0006(5)$ & $\pp0.0011(3)$ &  $25762$ \\
&&$2.00  $ & $0.1319400$& $\pp0.0043(4)$ & $-0.0029(2)$ &    $43986$ \\
\midrule
\multirow{3}{*}{3.8}&\multirow{3}{*}{12}
&  $1.20  $ & $0.1434300$& $\pp0.0095(5)$ & $\pp0.0012(3)$  &$ 14967$\\
& &$1.60  $ & $0.1376520$& $\pp0.0005(3)$ & $\pp0.0003(2)$  & $7000$ \\
& &$2.00  $ & $0.1319220$& $\pp0.00952(15)$ & $-0.00095(16)$& $12500$\\
\midrule
\multirow{4}{*}{4.3}&\multirow{4}{*}{8}
 &$1.00  $ & $0.1410350$& $\pp0.0071(13)$ & $\pp0.0045(5)$ & $6166$ \\
&&$1.30  $ & $0.1374570$& $\pp0.0053(5)$ & $\pp0.0015(4)$ & $6598$ \\
&&$1.60  $ & $0.1342250$& $-0.0035(5)$ & $-0.0010(4)$ & $5200$ \\
&&$1.90  $ & $0.1311370$& $-0.0144(7)$ & $-0.0042(4)$ & $6400$ \\
\midrule
\multirow{4}{*}{6.0}&\multirow{4}{*}{8}
&$1.00  $ & $0.1341200$& $-0.00058(20)$ & $\pp0.0029(3)$ & $6400$ \\
&&$1.20  $ & $0.1324830$& $\pp0.00517(16)$ & $\pp0.00053(14)$ & $7180$ \\
&&$1.40  $ & $0.1310350$& $\pp0.00394(17)$ & $-0.00174(13)$ & $6938$ \\
&&$1.60  $ & $0.1296320$& $\pp0.00118(11)$ & $-0.00397(11)$ & $7174$ \\
\bottomrule
\end{tabular}

\caption{\label{t:sims} Simulation parameters for the runs used 
in the $\csw$ determination as well as the resultant values for $aM$ and 
$a\Delta M$ calculated on those ensembles. The integrated molecular dynamics 
time of all replica for each ensemble is also given.  
For all of these runs the boundary fields are as specified in Eq.~\ref{e:bnd}.}
\end{table}

\begin{table}
\centering
\small
\newcommand{\pp}{\hphantom{-}}
\begin{tabular}{@{\extracolsep{0.3cm}}cccccccc}
\toprule
$\beta$ & $3.3$     &$3.4$      &$3.5$      &$3.601$    &$3.8$      & $4.3$      &$6.0$    \\ 
$\csw$  & $2.13(5)$ & $1.96(4)$ & $1.90(3)$ & $1.78(3)$ &$1.61(2)$  & $1.43(2)$  &$1.213(9)$ \\
\bottomrule
\end{tabular}

\caption{\label{t:csw} Values of the optimal $\csw$ for which $a\Delta M=a\Delta
M^{(0)}$. Each value is from a linear interpolation of $L/a=8$ data  
at a fixed value of $\beta$.}
\end{table}

\subsection{Analysis}\label{s:ana}

For each value of the coupling, we have to find the value of $\csw$ and
the quark mass for which the improvement condition is satisfied. It has
already been shown in the past that the condition  $M=0$ does not have
to be achieved to a very high accuracy~\cite{impr:csw_nf2,Tekin:2009kq}
and we therefore require $|aM| < 0.015$ for each individual point. We
then measure $a\Delta M$  for several choices of $\csw$ and interpolate
linearly to find the point with $\Delta M=\Delta M^{(0)}$. 

The measurements of the required fermionic correlation functions
(Eq.~\ref{e:cor}) are separated by one trajectory of length $\tau=2$.
Using the methods and software of Ref.~\cite{Wolff:2003sm}, we determine
the integrated autocorrelation times of the observables $aM$ and
$a\Delta M$ and find them to be at most around $8$ units of molecular
dynamics time such that all ensembles provide at least $500$ independent
measurements and thus a reliable determination of the errors. This is
confirmed by the normal distribution of mean values on single replica. 

Also, by studying the observables defined through the Wilson
flow~\cite{Luscher:2010iy} ---  the action density constructed from
smoothed links and the topological charge --- we ensure that field space
is sufficiently sampled in all simulations. In particular the
topological charge is known to cause problems in the continuum
limit~\cite{Schaefer:2010hu}.  However, since we keep $L/a$ fixed
instead of $L$, the contribution of sectors of non-zero topological
charge diminishes rapidly as $a\to0$.

The integrated autocorrelation times of the smoothed action and the
topological charge at various smoothing ranges do not  exceed $100$
units of molecular dynamic time in any of our simulations. We therefore
conclude that also for these observables configuration space is sampled
sufficiently.

\subsection{$\csw$ for $L/a = 8$}\label{s:l8}

\begin{figure}
\begin{center}
\includegraphics[width=0.5\textwidth,angle=-90]{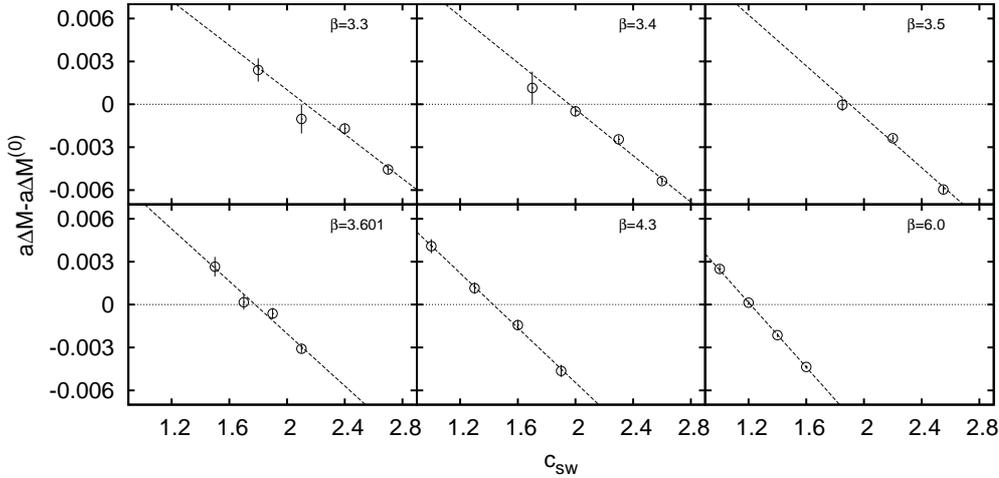}
\end{center}

\caption{\label{f:dm_v_csw} The improvement condition $\Delta M$ for several 
values of $\csw$ at each value of $\beta$. Linear fits to the improvement 
condition as a function of $\csw$ at fixed $\beta$ are also shown. The resultant 
values of $\csw$ which minimize the improvement condition are given in
Table~\ref{t:csw}.} 
\end{figure}

\begin{figure}
\begin{center}
\includegraphics[width=0.5\textwidth,angle=-90]{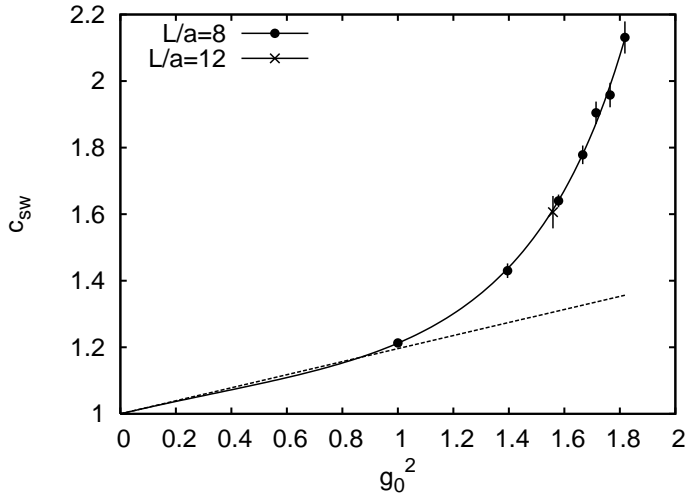}
\end{center}
\caption{\label{f:csw}Calculated values for $\csw$ together with the
   interpolating function represented by the solid line. The dashed line
is given by one-loop perturbation theory. To quantify finite volume
effects, a value from  simulations at $23\times12^3$ is given, with the
value for $g_0^2$ slightly shifted to the left for clarity.} 
\end{figure}

The results for the improvement condition as a function of $\csw$ as well 
as the resultant linear interpolations are shown in Fig.~\ref{f:dm_v_csw} for 
six values of $\beta$. The level of statistics is such that 
$\csw$ is determined with better than 3\% precision in all cases. 
The optimal $\csw$ values at fixed $\beta$ are 
collected in Tab.~\ref{t:csw} and shown in Fig.~\ref{f:csw} together with the 
best-fit interpolating curve  
\begin{align}
\csw(g_0^2) &= \frac{1-0.1921\, g_0^2-0.1378\, g_0^4+0.0717\,
   g_0^6}{1-0.3881\, g_0^2} \,.
\end{align}
We see that the fit describes the data well and that the data 
approaches the known 1-loop result at large $\beta$, which we 
use to constrain the interpolating curve.

\subsection{Finite volume check\label{s:fv}}

Although $\csw$ has typically been determined at fixed $L/a=8$, it is
interesting to assess the impact of this choice on the final result.
As stated above, it would actually be preferable to keep $L$
fixed when imposing the improvement condition.
Smaller volumes are
advantageous  as they are computationally cheaper and have a larger slope in 
$\Delta M$ vs. $\csw$, resulting in a clearer signal. Also, the problem of 
autocorrelations associated with freezing topological modes is much
reduced.
However, small volumes give rise to potentially large $\mathrm{O}(a)$ effects in $\csw$.

\begin{figure}
\begin{center}
\includegraphics[width=0.5\textwidth,angle=-90]{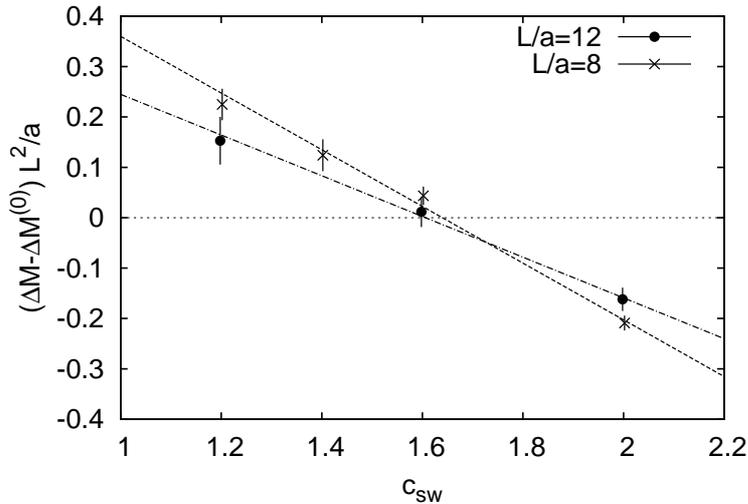}
\caption{\label{f:fv}A comparison of the improvement condition at $L/a=8$
and $L/a=12$ for $\beta=3.8$. The values of $\csw$ are slightly displaced 
for better clarity.} 
\end{center}
\end{figure}

We detail here a single check at $\beta=3.8$ between $L/a=8$ and
$L/a=12$. The result is shown in Fig.~\ref{f:fv}. Within the statistical
accuracies, the two volumes give the same value of $\csw$. With
$\csw=1.64(2)$ for $L/a=8$ and $1.61(5)$ for $L/a=12$, one can conclude
that at least up to that lattice spacing, the smaller lattices are a
good choice for the determination of $\csw$.  The point at $L/a=12$ also
does not deviate by more than one standard deviation from the
interpolating curve plotted in Fig.~\ref{f:csw}.  Therefore, given our
statistical accuracy, the systematic effects from fixing $L/a=8$ do not
appear to be significant.

\subsection{Preliminary scale determination}\label{s:scale}

To estimate the lattice spacing we compute a  quantity with
our discretization as well as the discretization used by PACS-CS, where the scale is
known from 
large volume simulations using the mass of the $\Omega$ baryon~\cite{Aoki:2008sm}. 
For comparison we use the coupling defined in 
Ref.~\cite{Fritzsch:2013je}, except with periodic spatial boundary conditions 
for the fermion fields, i.e. with $\theta=0$. This coupling is a renormalized 
quantity and at $M=0$ depends only on $L$, up to scaling violations.  
This lattice size $L$ at a given value of the coupling serves as the
dimensionful quantity to set the physical scale.

Our results for this coupling are tabulated in Table~\ref{t:scale}. 
In both discretizations we expect cutoff effects of order 
$\mathrm{O}(ag_0^2)$ as boundary improvement for the gauge fields is 
implemented at tree level only. Furthermore, boundary improvement for the 
fermion fields is implemented 
only at 1-loop, resulting in additional $\mathrm{O}(ag_0^4)$ effects. 
We see that the $\beta=3.3$ result for the coupling in our discretization lies 
above the value at $a=0.09\mathrm{fm}$, while the $\beta=3.4$ is below. This 
suggests that the $\beta$ corresponding to $a=0.09\mathrm{fm}$ in our 
discretization is in the range $[3.3,3.4]$.  

\begin{table}
\begin{center}
\begin{tabular}{ccccccccc}
\toprule
$L/a$ & $\beta$ & $c_1$ & $\csw$ & $c_F$ & $\kappa$ & $a(\mathrm{fm})$ & 
 $\bar{g}^2(L)$ & MDU \\
\midrule
$8$ & $1.9$ & $-0.331$  & $1.715$ & $0.972168$ & $0.1377$  & $0.090$  & $6.829(26)$ & $45614$  \\
\midrule
$8$ & $3.3$ & $-1/12$  & $2.127114$ & $0.970424$ & $0.137017$  & --  & $7.381(72)$ & $21904$  \\
$8$ & $3.4$ & $-1/12$  & $1.986246$ & $0.971294$ & $0.137553$  & --  & $6.225(21)$ & $38408$  \\
\bottomrule
\end{tabular}
\caption{\label{t:scale} $N_f = 3$ results for the Wilson flow coupling 
$\bar{g}^2(L)$ from Ref.~\protect\cite{Fritzsch:2013je} using the Iwasaki gauge action 
and our discretization. We take the parameters and the
scale determination from Ref.~\protect\cite{Aoki:2008sm}, while $\kappa$
was taken from Ref.~\protect\cite{Luscher:2012av} and found to give 
$aM\approx 0$. For these simulations only, we set $T=L$ with boundary fields $C_k = C_k' = 0$.}
\end{center}
\end{table}

\section{Conclusions}
\label{sec:conc}

In this paper we have determined $\csw$ for $N_f=3$ lattice QCD with  
the tree-level Symanzik-improved gauge action. The result 
of our determination is the interpolation formula
\begin{align}
\csw(g_0^2) &= \frac{1-0.1921\, g_0^2-0.1378\, g_0^4+0.0717\,
   g_0^6}{1-0.3881\, g_0^2} \,,
\end{align}
which may be taken as a definition of the lattice action. While our 
determination was performed at fixed $L/a=8$, we performed a finite volume 
check at $\beta=3.8$ and found no significant change in $\csw$.   

In addition to a determination of $\csw$, we have calculated a
renormalized $L$-dependent coupling at $\beta=3.3$ and $3.4$ and
compared with an alternative $N_f =3$ discretization. This determination
suggests that the bare coupling corresponding to $a \approx
0.09\mathrm{fm}$ in our discretization is located in the interval $\beta
\in [3.3,3.4]$, indicating that our $\csw$ determination spans the range
of desired lattice spacings.  

\section*{Acknowledgments}
It is a pleasure to thank Rainer Sommer and Martin L\"uscher for
numerous and  very helpful discussions and Rainer Sommer for a careful
reading of an earlier version of this manuscript. We are also grateful
to Hubert Simma and Christian Wittemeier  who kindly provided us with
their measurement code.  The simulations have been done on the thqcd2
installation at CERN. We want to thank the IT department for essential
support.

\providecommand{\href}[2]{#2}\begingroup\raggedright\endgroup

\end{document}